\documentclass[onecolumn,superscriptaddress,showpacs]{revtex4-1}
\usepackage{enumerate,appendix}
\usepackage{amsmath, amsthm, amssymb,commath,braket}
\usepackage{color,calc,graphicx}
\usepackage[colorlinks]{hyperref}

\usepackage{graphicx}% Include figure files
\usepackage{dcolumn}% Align table columns on decimal point
\usepackage{bm}% bold math
\usepackage[all]{xy}
\usepackage{indentfirst}
\usepackage{amsmath}
\usepackage{bbm}
\usepackage{color}
\usepackage{multirow}
\usepackage{mathrsfs}
\usepackage{euscript}
\usepackage{amssymb}

\newcommand{\va}{\vec{a}}
\newcommand{\vx}{\vec{x}}
\newcommand{\vu}{\vec{u}}
\newcommand{\vA}{\vec{A}}

\newcommand{\dt}{\diamondsuit}

\newcommand{\oI}{\overline{I}}

\newcommand{\vsA}{\vec{\textsf{A}}}

\newcommand{\sA}{\textsf{A}}
\newcommand{\sB}{\textsf{B}}
\newcommand{\sC}{\textsf{C}}

\newcommand{\cH}{\mathcal{H}}
\newcommand{\cW}{\mathcal{W}}

\newcommand{\cN}{\mathcal{N}}

\newtheorem{definition}{Definition}

\begin{document}
%%%%%%%%%%%%%%%%%%%%%%%%%%%%%%%%%%%%%%%%%%%%%%%%%%%%%%%%%%%%%%%%%%%%%%%%%%%%%
%%%%%%%%%%%%%%%%%%%%%%%%%%%%%%%%%%%%%%%%%%%%%%%%%%%%%%%%%%%%%%%%%%%%%%%%%%%%%
\title{Verifying Hierarchic Multipartite and Network Nonlocalities with a Unified Method}
%%%%%%%%%%%%%%%%%%%%%%%%%%%%%%%%%%%%%%%%%%%%%%%%%%%%%%%%%%%%%%%%%%%%%%%%%%%%%
\author{Ming-Xing Luo}
\email{mxluo@swjtu.edu.cn}
\affiliation{School of Information Science and Technology, Southwest Jiaotong University, Chengdu 610031, China}
\affiliation{CAS Center for Excellence in Quantum Information and Quantum Physics, Hefei, 230026, China}

\author{Shao-Ming Fei}

\affiliation{School of Mathematical Sciences, Capital Normal University, Beijing 100048, China}
\affiliation{Max-Planck-Institute for Mathematics in the Sciences, 04103 Leipzig, Germany}

\begin{abstract}
The multipartite nonlocality provides deep insights into the fundamental feature of quantum mechanics and guarantees different degrees of cryptography security for potential applications in the quantum internet. Verifying multipartite nonlocal correlations is a difficult task. We propose a unified approach that encompasses all the quantum characteristics of the multipartite correlated system beyond from fully separable to biseparable no-signaling correlations. We offer a straightforward method to verify general systems by lifting partial nonlocal correlations. This allows to construct a chained Bell inequality, facilitating the unified verification of hierarchic multipartite nonlocalities. We finally apply the lifting method to verify the correlations derived from quantum networks.
\end{abstract}
%%%%%%%%%%%%%%%%%%%%%%%%%%%%%%%%%%%%%%%%%%%%%%%%%%%%%%%%%%%%%%%%%%%%%%%%%%%%%
%%%%%%%%%%%%%%%%%%%%%%%%%%%%%%%%%%%%%%%%%%%%%%%%%%%%%%%%%%%%%%%%%%%%%%%%%%%%%
%%%%%%%%%%%%%%%%%%%%%%%%%%%%%%%%%%%%%%%%%%%%%%%%%%%%%%%%%%%%%%%%%%%%%%%%%%%%%
\maketitle
\section{Introduction}

Bell's theorem, originated from John Bell's profound insights \cite{Bell}, is a crucial method to investigate the intricate aspects of quantum mechanics. The violation of Bell inequality enables us to uncover the nonlocality of bipartite quantum systems and surpass the limitations imposed by local realism, ultimately addressing the Einstein-Podolsky-Rosen (EPR) argument \cite{EPR}. Employing Bell inequalities provides an efficient method to ascertain the existence of entanglement between space-separated quantum systems \cite{CHSH,Gisin,GP}. Bell's theorem encompasses both the conceptual significance and practical implications, thereby stimulating further exploration into quantum phenomena and potential applications \cite{Brunner}.

The Bell nonlocality can be extended to characterize the nonlocal correlations arising from multipartite systems. Interestingly, some multiple systems allow for the generation of global correlations that cannot be simulated using partially entangled systems. The so-called genuine multipartite nonlocality can be traced back to the seminal works of Svetlichny \cite{Sy} and Greenberger, Horne and Zeilinger (GHZ) \cite{GHZ}. Genuine entanglement is not limited to pairs of particles but can exist among multiple particles, leading to stronger violations of local realism \cite{PR,JC}, even with general postselection \cite{GPS}. Under the local decomposition this model enables to feature any underlying networks consisting of some nonlocal resources \cite{CWR,BP}. Exploring these genuine nonlocal correlations may significantly enhance the understanding of the intricate nature of entanglement. Nevertheless, verifying multipartite entanglement is a difficult task for general systems \cite{Gurvits}.

Several well-known methods have been proposed to address this challenge. The Bell inequalities provide a device-independent method for multipartite scenarios, such as the Svetlichny inequality \cite{Sy,JC,BBGL}, Mermin inequality \cite{Mermin,Chen2004,Fei} and Hardy inequality \cite{Hardy,Chen}. Other involve the utilization of entanglement witnesses \cite{TG,GW,ZD,FR,HHH} or EPR-steering \cite{WJD07,He,YSQ}. These ways generally exhibit different degrees of  multipartite correlations \cite{SG,Zuko,CL,RD}. Remarkably, a recent result shows that all of these genuine nonlocalities are equivalent in the presence of multiple identical entangled isolated systems \cite{Luo2023}. Verifying these correlations through a unified method remains an open problem, which is very important in experimental implementations with many-body systems.

\begin{figure}
\begin{center}
\includegraphics[width=0.49\textwidth]{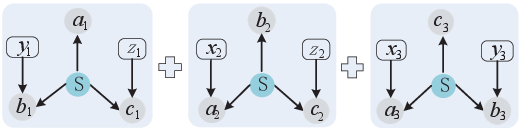}
\end{center}
\caption{\small (Color online) (a) Chained tripartite Bell experiments. Each node in the directed acyclic graphs (DAGs) represents a variable that denotes the type of measurements ($x_i, y_i, z_i$), measurement outcomes ($a_i, b_i, c_i$), or source $S$. Each directed edge of two nodes encodes their causal dependence relation. The experiment is to jointly feature three postselected  bipartite probability correlations $\{P(b_1c_1|y_1z_1;a_1)\}$, $\{P(a_2c_2|x_2z_2;b_2)\}$ and $\{P(a_3b_3|x_3y_3;c_3)\}$.
}
\label{fig00}
\end{figure}

In this work, we propose a unified way to encompass separable correlations, even in the absence of the no-signaling principle \cite{PR}. Our model enables the verification of hierarchic multipartite nonlocal correlations by violating a unified Bell inequality \cite{Sy,Gal,Ban}. An example of tripartite Bell test is shown in Fig.\ref{fig00} under multiple measurement settings. While any activated bipartite correlations in the biseparable model \cite{Sy} exhibits a maximal bound of three CHSH operators at 8 \cite{CHSH}, the local measurements on specific tripartite state can activate bipartite quantum correlations which yield a maximal violation. We extend this approach to verify hierarchic nonlocalities of general multipartite states. Interestingly, it provides a unified method to verify the correlations from  networks \cite{Alex,LuoMX2023,TPLR}. We finally witness the genuine entanglement by lifting some entanglement witnesses on low-dimensional states \cite{Toth2005}.

\section{Verifying multipartite nonlocalties}

\subsection{Hierarchic correlations from one entanglement}

Consider an $n$-partite Bell test with a source $S$ that distributes states to space-like separated observers $\sA_1, \cdots, \sA_n$ \cite{Bell}. The measurement outcome $a_i$ of $\sA_i$ depends on local shares and the type of measurement denoted by $x_i$. The joint distribution of all outcomes, conditional on measurement settings, is considered biseparable if it can be decomposed as follows:
\begin{eqnarray}
P_{bs}(\va|\vx)
=\sum_{I,\oI}p_{I,\oI}P(\va_I|\vx_I) P(\va_{\oI}|\vx_{\oI}),
\label{eqn-2}
\end{eqnarray}
where $\vu:=(u_1,\cdots, u_n)$ for $u=a$ or $x$, $\vu_{J}:=(u_i,i\in J)$, $a_i$ and $x_i$ denote inputs and outcomes per party, $I$ and $\oI$ are bipartition of $\{1, \cdots, n\}$, and $\{p_{I,\oI}\}$ is a probability distribution over all bipartitions. $P(\va_I|\vx_I)$ denote the joint distribution of the outcomes $\va_I$ conditional on the measurement settings $\vx_I$, and similar notation is used for $P(\va_{\oI}|\vx_{\oI})$.

The correlation (\ref{eqn-2}) is \textit{NS (no-signalling) biseparable} if both $P(\va_I|\vx_I)$ and $P(\va_{\oI}|\vx_{\oI})$ are NS correlations \cite{PR,Gal,Ban}, that is, $p(a_i|x_i,x_j)=p(a_i|x_i)$ for any $i$ and $j\not=i$. This kind of correlations are regarded as \textit{nonlocal biseparable} if all the involved partial correlations $P(\va_I|\vx_I)$ and $P(\va_{\oI}|\vx_{\oI})$ are generated from quantum states. Inspired by the EPR-steering \cite{EPR,He}, define the
\textit{EPR biseparable} if all the partial correlations $P(\va_I|\vx_I)$ and $P(\va_{\oI}|\vx_{\oI})$ are any EPR-steering correlations. The \textit{quantum biseparable} is defined with quantum states \cite{HHH}, i.e., the quantum state cannot be decomposed into the mixturing of biseparable states. \textit{Bell fully separable} correlations refers the decomposition with classical ones from hidden variables \cite{Bell}. This leads hierarchic multipartite correlations as follows:

\begin{definition}
The joint correlation $P(\va|\vx)$ is classified as
\begin{itemize}
\item{} genuine multipartite nonlocal (GMN$^*$) if it is not biseparable;

\item{} genuine multipartite nonlocal (GMN) if it is not NS biseparable;

\item{} $k$-level genuine multipartite nonlocal (GMN$^{(k)}$) if it is not biseparable and local joint distribution of no more than $k$ parties is NS;

 \item{} genuine multipartite quantum nonlocal (GMQN) if it is not biseparable in terms of quantum correlations;

\item{} genuine multipartite EPR-steering nonlocal (GEPRN) if it is not EPR-steering biseparable;

\item{} genuine multipartite entanglement (GME) if it is not biseparable in terms of quantum states;

\item{} multipartite Bell nonlocal (MBN) if it is not Bell fully separable.
\end{itemize}

\label{def-1}
\end{definition}

According to Definition \ref{def-1}, the GMN provides the strong nonlocality of multipartite correlations under the NS principle, which can be verified by violating specific Bell inequalities \cite{Mermin,Sy}. However, in some applications such as quantum secret sharing \cite{Hillery,Cleve,Moreno,Luo2022}, a subset of participants may cooperate to recover others' outcomes. This requires strong nonlocal correlations to guarantee the security by ruling out any biseparable decomposition (\ref{eqn-2}) without the NS restrictions on all partial correlations. Similar scenarios may require to characterize multipartite correlations. One is GMN$^{(k)}$ under the assumption that no more than $k$ number of parties allow to perform joint local measurements. The MBN is the weakest form of the nonlocality that can be derived from local entangled states \cite{CHSH,Mermin}.

Note that Definition 1 has not considered all the local models for multipartite entangled states. One example is to rule out network decomposition \cite{CWR,Luo2021}. The other is to consider quantum preparation \cite{LuoF2023}. Interestingly, given multiple identical and independent entangled states, recent results provide the first efficient method to verify GMN of all entangled pure states \cite{Luo2023}. This implies the equivalence of GMN, GMQN, GEPRN, and GME. But, verifying multipartite correlations of general states is an NP-hard problem \cite{Gurvits}. Our goal is to develop a unified method to explore hierarchical correlations. This depends on the standard assumption for each party, i.e., without distinguishing local operations \cite{CWR,BP}.

\subsection{Verifying multipartite nonlocalties with lifting partial correlations}

In the context of NS theories, correlations between different systems can be generated by using NS source such as PR-box \cite{PR,JC,Barrett}. Performing experiments on NS sources can then build the NS correlations under specific hypotheses that are satisfied by both classical variable models and quantum mechanics \cite{Luo2023}. This allows to generate any biseparable correlations (\ref{eqn-2}).

Consider the verification of GMN for a general $m$-partite state $\rho$ on Hilbert space $\otimes_{i=1}^{n+1}\mathcal{H}_{A_i}$. Verifying GMN requires to build an $n$-partite Bell inequality shown as
\begin{eqnarray}
\mathcal{B}:=\sum_{x_1,\cdots, x_n}\alpha_{x_1\cdots x_n}
A_{x_1}^{(1)}\cdots A_{x_n}^{(n)}\leq c,
\label{eqnn2}
\end{eqnarray}
where $\{A_{x_i}^{(i)}, \forall x_i\}$ are measurement operators of the $i$-th party satisfying $|A_{x_i}^{(i)}|\leq 1$, and $\alpha_{x_1\cdots x_n}$ are constants. Define an $m$-partite Bell test that contains $n+1$ post-selections. In the $i$-th sub-test, the observer $\textsf{A}_i$ has one measurement setting with outcome $a_i^{0}$, while others have multiple measurement settings. Denote all observers except for $\textsf{A}_i$ by $\vA_i$. A main result is the following chain Bell inequality for any biseparable source:
\begin{eqnarray}
\sum_{i=1}^{n+1}\mathcal{B}(\vA_i)\leq c_{ns}+nc,
\label{eqn-5}
\end{eqnarray}
where $c_{ns}$ denotes the upper bound of $\mathcal{B}$ in terms of the NS correlation. The violation of the inequality (\ref{eqn-5}) provides a new method to verify $n+1$-partite nonlocality.

\textbf{Proof of Inequality (\ref{eqn-5})}. Consider a Bell-type test in the biseparable no-signaling (NS) model, where all observers
$\sA_i$ share a biseparable NS source that can be used to generate the biseparable distribution (\ref{eqn-2}). For any bipartition of $\{\sA_i\}$ and $\{\sA_j, j\not=i\}$, we rewrite it into
\begin{eqnarray}
\dt_{A_1\cdots{}A_{n+1}}
=\sum_{i=1}^{n+1}p_{i}\dt_{A_i}\bot \dt_{\vA_{i}},
\label{NSsourcen}
\end{eqnarray}
where $\dt_{\vA_{i}}$ denotes a NS source shared by all parties in $\vA_{i}$, and $\vA_{i}$ denotes all parties except for $\textsf{A}_i$ (see SI Section A). Denote $\{M^{a_k^0}\}$ as local measurements of $\sA_k$ with $k\in \{1,\cdots, n+1\}$, and $\{N_{a_s^{_{(k)}}|x_s^{_{(k)}}}\}$ as local measurements of $\sA_s$ with $s\not=k$. The main goal is to represent joint distribution of all $n$ observers conditional on local outcomes $a_k^0$, $k=1, \cdots, n+1$. The following proof is based on Lemma 1  (See Section B \cite{SI}).

\textbf{Lemma 1}. For a given $k$ the joint distribution of the outcomes $a_s^{_{(k)}}$ with $s\not=k$, conditional on the outcomes $a_k^0$, is given by
\begin{eqnarray}
&P_{a_k^0}(\va_{k}^{_{(k)}}|\vx_{k}^{_{(k)}})=p_kP_{ns}(\va_{k}^{_{(k)}}|\vx_{k}^{_{(k)}})
\nonumber
\\
&+(1-p_k)P_{bs}(\va_{k}^{_{(k)}}|\vx_{k}^{_{(k)}}),
\label{SA1}
\end{eqnarray}
where $\va_{i}^{_{(k)}}=(a_j^{_{(k)}},j\not=i)$, $\vx_{i}^{_{(k)}}=(x_j^{_{(k)}},j\not=i)$, $P_{ns}(\va_{k}^{_{(k)}}|\vx_{k}^{_{(k)}})$ can be any NS bipartite distribution and $P_{bs}(\va_{k}^{_{(k)}}|\vx_{k}^{_{(k)}})$ is biseparable distribution.

Consider a Bell test with the biseparable NS source. The observer $\sA_k$ performs local measurements $\{M^{a_k^0}\}$ on the shared state while each observer $\sA_s$ for any $s\not=k$ performs local measurements $\{M^{a_s}_{x_s}\}$ on the shared states. The goal is to explore $n$-partite correlation $P(\va_k|\vx_k)$ conditional on the local outcome $a_k^0$. Consider any linear Bell correlator $\mathcal{B}$ defined in Eq.(\ref{eqnn2}) for any $n$-partite biseparable correlation. From Lemma 1 it implies for any outcomes $a_1^0,\cdots, a_{n+1}^0$ that
\begin{eqnarray}
\sum_{k=1}^{n+1}\mathcal{B}|_{\{P(\va_k^{_{(k)}}|\vx_{k}^{_{(k)}};a_k^0)\}}&=& \sum_{k=1}^{n+1}p_k\mathcal{B}|_{\{P_{ns}(\va_k^{_{(k)}}|\vx_k^{_{(k)}})\}}
\nonumber
\\
&&+\sum_{k=1}^{n+1}(1-p_k)\mathcal{B}|_{\{P_{bs}(\va_k^{_{(k)}}|\vx_{k}^{_{(k)}};a_k^0)\}}
\nonumber
\\
&\leq & c_{ns}+n c,
\label{SA9}
\end{eqnarray}
where $c_{ns}$ denotes the upper bound of Bell operator (\ref{eqnn2}) with respect to the NS correlations. This has completed the proof. $\Box$

To illustrate our main idea, we verify tripartite correlations. Take $\mathcal{B}$ as the bipartite CHSH operator \cite{CHSH}, i.e., $\mathcal{B}(X,Y)=\sum_{i,j=0,1}(-1)^{i\cdot{}j}X_iY_j$ with the correlator $X_iY_j=\sum_{x,y=0,1}(-1)^{x+y}P(a,b|x,y)$. We get a chain Bell inequality (Section C \cite{SI}):
\begin{eqnarray}
\Delta_3&:=&\sum_{i=1}^3\mathcal{B}(A^{(i)},B^{(i)})\leq
\left\{
\begin{array}{lll}
6, & \textsf{FS};
\\
4+2\sqrt{2}, & \textsf{BQS};
\\
8, & \textsf{BS};
\\
6\sqrt{2}, & \textsf{Q};
\\
12, & \textsf{NS}/\textsf{GC},
\end{array}
\right.
\label{tripartite}
\end{eqnarray}
where \textsf{FS} denotes full separable correlations, \textsf{BQS} denotes biseparable quantum states, \textsf{BS} denotes biseparable correlations, \textsf{Q} denotes quantum correlations, \textsf{NS} denotes NS correlations and \textsf{GC} denotes general correlations. This inequality is different from previous monogamy inequality \cite{Yang2024} and provides the first a unified method to verify hierarchic tripartite correlations.

\textit{Example 1}. Consider a generalized GHZ state \cite{GHZ} as:
 \begin{eqnarray}
\ket{\Phi}_{ABC}=\cos\theta\ket{000}+\sin\theta\ket{111},
\end{eqnarray}
where $\theta\in (0,\frac{\pi}{2})$. Define $M^{a_1^0},M^{a_2^0}, M^{a_3^0}\in \{\sin\theta\ket{0}+\cos\theta\ket{1}, \cos\theta\ket{0}-\sin\theta\ket{1}\}$, $M_{a_2}^{_{(1)}},M_{a_1}^{_{(2)}}, M_{a_1}^{_{(3)}}\in \{\sigma_z,\sigma_x\}$ and $M_{a_3}^{_{(1)}}, M_{a_3}^{_{(2)}},  M_{a_2}^{_{(3)}}\in \{(\sigma_z\pm \sigma_x)/\sqrt{2}\}$. For the outcome $a_1^0=0$, $a_2^0=0$, or $a_3^0=0$, the resultant is an EPR state $(\ket{00}+\ket{11})/\sqrt{2}$ \cite{EPR}. The following inequality holds for quantum correlations as
\begin{eqnarray}
\sum_{k=1}^3\textsf{CHSH}(P_{a_k^0}(\va_k^{_{(1)}}|\vx_k^{_{(1)}}))= 6\sqrt{2}>8
\end{eqnarray}
which violates the inequality (\ref{tripartite}) for any $\theta\in (0,\frac{\pi}{2})$. This means under the post-selection there are local observables for all observers such that bipartite correlations violate the inequality (\ref{tripartite}) with the maximal bound $6\sqrt{2}$. The result can be extended for high-dimensional GHZ states.

\textbf{Example 2}. Consider a generalized W state \cite{Dur}:
\begin{eqnarray}
&|W\rangle=\cos\theta_1\cos\theta_2\ket{001}+\cos\theta_1\sin\theta_2\ket{010}
\nonumber\\
&+\sin\theta_1\ket{100},
\end{eqnarray}
where $\theta_1,\theta_2\in (0,\frac{\pi}{2})$. Define $M^{a_1^0}, M^{a_2^0}, M^{a_2^0}\in \{\ket{0}$, $\ket{1}\}$. For the outcome $a_1^0=0$, the reduced state of $A_2A_3$ is $|\phi\rangle=\cos\theta_2\ket{01}
+\sin\theta_2\ket{10}$. The maximal bound of CHSH operator \cite{CHSH} with respect to this state is given by $2\sqrt{1+\sin^22\theta_2}$. This implies there are local observables $M_{a_2}^{_{(1)}}$ and $M_{a_3}^{_{(1)}}$ such that quantum correlations satisfy
\begin{eqnarray}
\textsf{CHSH}(P_{a_1^0}(a_2^{_{(1)}}, a_3^{_{(1)}}|x_2^{_{(1)}}, x_3^{_{(1)}}))
=2\sqrt{1+\sin^22\theta_2}.
\label{S21}
\end{eqnarray}

Similarly, there are local observables $M_{a_1}^{_{(2)}}$ and $M_{a_3}^{_{(2)}}$ such that quantum correlations for the measurement outcome $a_2^0=0$ satisfy that
\begin{eqnarray}
&&\textsf{CHSH}(P_{a_2^0}(a_1^{_{(2)}},a_3^{_{(2)}}|x_1^{_{(2)}},x_3^{_{(2)}}))
\nonumber
\\
&=&2\sqrt{1+\frac{\sin^22\theta_1\cos^2\theta_2}{
\cos^2\theta_1\cos^2\theta_2+\sin^2\theta_1}}.
\label{S22}
\end{eqnarray}

\begin{figure}
\begin{center}
\includegraphics[width=0.5\textwidth]{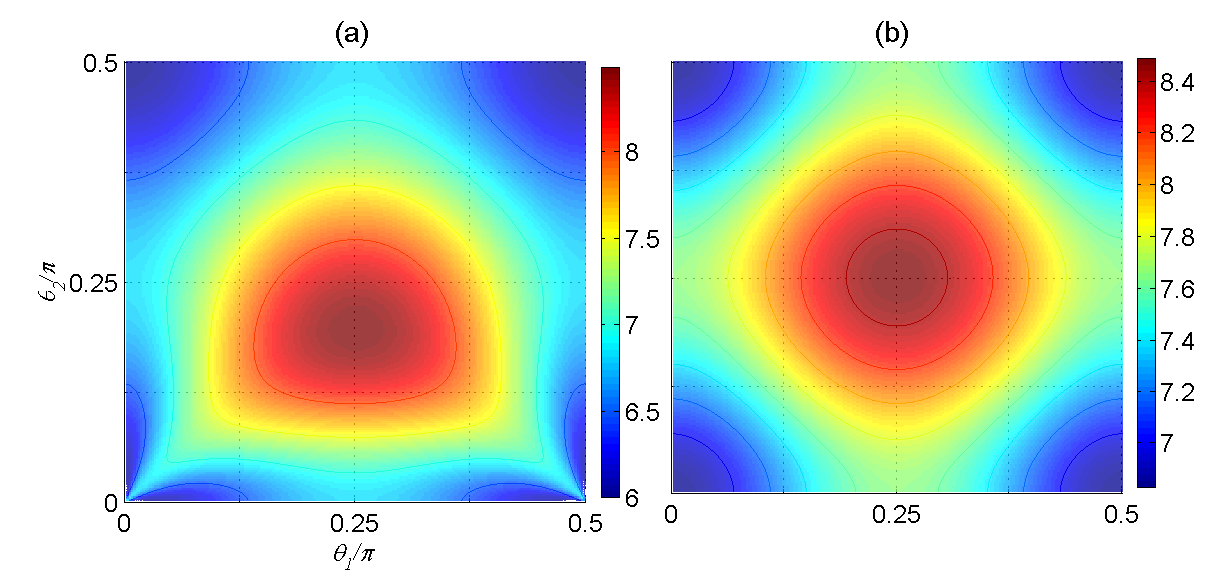}
\end{center}
\caption{\small (Color online) Schematic upper bounds of $\Delta_3$. (a) W state. (b) Chain network consisting of two generalized EPR states.
}
\label{fig-2}
\end{figure}

For the measurement outcome $a_3^0=0$, there are local observables $M_{a_1}^{_{(3)}}$ and $M_{a_2}^{_{(3)}}$ such that quantum correlations satisfy that
\begin{eqnarray}
&&\textsf{CHSH}(P_{a_3^0}(a_1^{_{(3)}},a_2^{_{(3)}}|x_1^{_{(3)}},x_2^{_{(3)}}))
\nonumber
\\
&=&2\sqrt{1+\frac{\sin^22\theta_1\sin^2\theta_2}{
\cos^2\theta_1\sin^2\theta_2+\sin^2\theta_1}}.
\label{S23}
\end{eqnarray}
Eqs.(\ref{S21}-\ref{S23}) imply the following inequality
\begin{eqnarray}
&&\sum_{k=1}^3\textsf{CHSH}(P_{a_k^0}(\va_k^{_{(k)}}|\vx_k^{_{(k)}}))
\nonumber
\\
&=&2\sqrt{1+\sin^22\theta_2}+
2\sqrt{1+\frac{\sin^22\theta_1\cos^2\theta_2}{
\cos^2\theta_1\cos^2\theta_2+\sin^2\theta_1}}
\nonumber
\\
&&+2\sqrt{1+\frac{\sin^22\theta_1\sin^2\theta_2}{
\cos^2\theta_1\sin^2\theta_2+\sin^2\theta_1}}
\end{eqnarray}
which violates the inequality (\ref{tripartite}) for some $\theta_i$'s. Figure \ref{fig-2}(a) shows the bound of Bell operator of $\Delta_3$ with tripartite correlations of a generalized W state.

\textbf{Example 3}. Consider a chain network consisting of two generalized EPR states \cite{EPR}: $|\phi_i\rangle=\cos\theta_i\ket{00}+\sin\theta_i\ket{11}$ with $\theta_i\in (0, \frac{\pi}{2})$, $i=1, 2$. Here, $\textsf{A}$ has the particle $1$, $\textsf{B}$ has the particles $2$ and $3$, and $\textsf{C}$ has the particle $4$. Define $M^{a_1^0},M^{a_3^0}\in \{\ket{0},\ket{1}\}$, $M^{a_2^0}\in \{\sin\varphi\ket{00}+\cos\varphi\ket{11}, \cos\varphi\ket{00}-\sin\varphi\ket{11}, \ket{01},\ket{10}\}$ with $\tan\varphi=1/(\tan\theta_1\tan\theta_2)$, $M_{a_2}^{_{(1)}},M_{a_1}^{_{(2)}}, M_{a_1}^{_{(3)}} \in \{\sigma_z$, $\sigma_x\}$, and $M_{a_3}^{_{(1)}}\in \{\cos\hat{\theta}_1\sigma_z\pm \sin\hat{\theta}_1\sigma_x\}$, $M_{a_3}^{_{(2)}}\in \{\cos\hat{\theta}_2\sigma_z\pm \sin\hat{\theta}_2\sigma_x\}$ and $M_{a_2}^{_{(3)}}\in \{\cos\hat{\theta}_3\sigma_z\pm \sin\hat{\theta}_3\sigma_x\}$. From the CHSH inequality it follows that
\begin{eqnarray}
&&\max_{\hat{\theta}_1}\textsf{CHSH}(P_{a_1^0}(a_2^{_{(1)}}, a_3^{_{(1)}}|x_2^{_{(1)}}, x_3^{_{(1)}}))
\nonumber
\\
&=&\max_{\hat{\theta}_1}\left\{
2\cos\hat{\theta}_1+2\sin \hat{\theta}_1\sin2\theta_1
\right\}
\nonumber
\\
&=&2\sqrt{1+\sin^2\theta_1}.
\end{eqnarray}
Similarly, we can prove that
\begin{eqnarray}
\max_{\theta_3,\hat{\theta}_3}
\textsf{CHSH}(P_{a_3^0}(a_1^{_{(3)}},a_2^{_{(3)}}|x_1^{_{(3)}},x_2^{_{(3)}}))=2\sqrt{1+\sin^2\theta_2}.
\end{eqnarray}
For the outcome $b^0=0$, the resultant is an EPR state. This implies that
\begin{eqnarray}
\max_{\hat{\theta}_2} \textsf{CHSH}(P_{a_2^0}(a_1^{_{(2)}},a_3^{_{(2)}}|x_1^{_{(2)}},x_3^{_{(2)}}))
=2\sqrt{2}.
\end{eqnarray}
So, we obtain the following inequality
\begin{eqnarray}
\sum_{k=1}^3\textsf{CHSH}(P_{a_k^0}(\va_k^{_{(k)}}|\vx_k^{_{(k)}}))
&=&2\sqrt{2}+2\sqrt{1+\sin^2\theta_1}
\nonumber
\\
&&+2\sqrt{1+\sin^2\theta_2}>8
\end{eqnarray}
if $\theta_i$ satisfy $\sqrt{1+\sin^2\theta_1}+\sqrt{1+\sin^2\theta_2}>4-\sqrt{2}$. This verifies the genuine tripartite nonlocality of chain networks by violating the inequality (\ref{tripartite}). Moreover, we can prove that $2\sqrt{2}+2\sqrt{1+\sin^2\theta_1}+2\sqrt{1+\sin^2\theta_2}>2\sqrt{2}+4$ for any
$\theta_i\in (0, \frac{\pi}{2})$. The genuine tripartite quantum nonlocality of any chain network can be verified by violating the inequality (\ref{tripartite}). Figure \ref{fig-2}(b) shows hierarchic correlations from tripartite chain networks.

Now, we consider the $n$-partite Svetlichny-Mermin operator $\mathcal{B}$ \cite{Sy,Mermin}. We prove a chain Bell inequality as (Section D \cite{SI}):
\begin{eqnarray}
\sum_{i=1}^{n+1}\mathcal{B}(\vec{A}^{(i)})\leq
\left\{
\begin{array}{lll}
(n+1)2^{n-1}, & \textsf{FS};
\\
(n+\sqrt{2})2^{n-1}, & \textsf{BQS};
\\
(n+2)2^{n-1}, & \textsf{BS};
\\
(n+1)2^{n-0.5}, & \textsf{Q};
\\
(n+1)2^{n}, & \textsf{NS}/\textsf{GC}
\end{array}
\right.
\label{npartite}
\end{eqnarray}
where \textsf{FS}, \textsf{BQS}, \textsf{BS}, \textsf{Q} and \textsf{NS}/\textsf{GC} are defined in Eq.(\ref{tripartite}).

Given an $n$-qubit state $\rho$, we will prove the maximal quantum bound of the Svetlichny-Mermin operator as follows. Define each observable as the form $A_{x_j=0}=\vec{s}^{(j)}
\cdot \vec{\sigma}=\Sigma_{k=1}^3 s_{k}^{(j)}\sigma_{k}$, and $A_{x_j=1}=\vec{t}^{(j)}
\cdot \vec{\sigma}=\Sigma_{k=1}^3 t_{k}^{(j)}\sigma_{k}$, for $j=1, \cdots, n$, where $\vec{\sigma}=(\sigma_{1},\sigma_{2},\sigma_{3})$ with Pauli matrices $\sigma_{i}$, $i=1,2,3$, and $\vec{s}$ and $\vec{t}$ are three-dimensional real unit vectors. For any $n$-partite state $\rho$ the maximal quantum bound of the Svetlichny operator satisfies
\begin{equation}
\mathcal{Q}(\mathcal{B}_n)={\rm max}\left |\langle \mathcal{B}_n \rangle _{\rho}\right |\leq 2^{n-1}{\lambda_{max}},
\label{Qbound}
\end{equation}
where $\lambda_{max}$ is the maximal singular value of the matrix $X$ defined by $X_{i_1,i_2\cdots i_n}= {\rm Tr}[\rho(\sigma_{i_1}\otimes \cdots \otimes \sigma_{i_n})]$, $i_1, \cdots, i_n=1, 2, 3$, which takes over all quantum states that are locally unitary equivalent to $\rho$. Equivalently, $\lambda_{1}$ is the maximal singular value of the matrix $M=(M_{i_1,\vec{i}_1})$ with $M_{i_1,\vec{i}_1}={\rm Tr}[\rho(\sigma _{i_1}\otimes\cdots \otimes \sigma _{i_n})]$.

In fact, from the double cover relationship between the special unitary group SU(2) and the special orthogonal group SO(3) \cite{Schlienz,Sun}, it follows taht $U\sigma_{_{i}} U^{\dagger}=\sum_{j=1}^{3}O_{ij}\sigma_{j}$, where $U$ is a given unitary operator and the matrix $O$ with entries $O_{ij}$ belongs to SO(3). We obtain that
\begin{eqnarray}
M_{i_1,\vec{i}_1}
&=&{\rm Tr}[\rho(U_1\sigma_{i_1}U_1^\dag\otimes \cdots \otimes U_n\sigma_{i_n}U_n^\dag)]
\nonumber \\
 &=&{\rm Tr}\left [ \rho (U_1 \Sigma _{A_1}U_1^{\dagger}\sigma_{i}U \Sigma _{A}U^{\dagger}\right ]
\nonumber \\
&=&\sum_{j_1,\cdots, j_n}O_{i_1j_1}^{(1)}\cdots{}O_{i_nj_n}^{(n)}{\rm Tr}\left [ \varrho (\sigma_{j_1} \otimes \cdots\otimes \sigma_{j_n})  \right ]\nonumber \\
&=&\left [ O_{1}X\left ( O_{2}^{T}\otimes \cdots \otimes O_{n}^{T} \right ) \right ]_{i_1|i_2\cdots i_n}.
\end{eqnarray}
This implies that $M=[O_{1}X( O_{2}^{T}\otimes \cdots \otimes O_{n}^{T})]_{i_1|i_2\cdots i_n}$, and
\begin{eqnarray}
M ^{\dag}M &=&( O_{2}\otimes \cdots \otimes O_{n}) X^\dag  O_{1}^\dag {}O_1 X ( O_{2}\otimes \cdots \otimes O_{n})^\dag
\nonumber \\
&=&( O_{2}\otimes \cdots \otimes O_{n}) X^\dag X( O_{2}\otimes \cdots \otimes O_{n})^\dag.
\end{eqnarray}
From the orthogonality of the operator $O_{2}\otimes \cdots \otimes O_{n}$, we obtain that $MM^\dag$ has the same eigenvalues as $X^{\dag}X$. Hence, $M$ and $X$ have the same singular values. Let $\vec{v}$ be a $3^{n-1}$-dimensional singular vector of the matrix $X$. Then $(O_{2}\otimes \cdots \otimes O_{n})\vec{v}$ is a $2^{n-1}$-dimensional singular vector of the matrix $M$. From Lemma \cite{Li}, i.e., for any vectors $\vec{x}\in R^m$ and $\vec{y}\in R^n$, it implies that
\begin{eqnarray}
|\vec{x}^T A \vec{y}|\leq \lambda_{max}|\vec{x}| \cdot |\vec{y}|,
\end{eqnarray}
where $\lambda_{max}$ is the largest singular value of the matrix $A$ of size $m \times n$. The equality holds when $\vx$ and $\vec{y}$ are the corresponding singular vectors of $A$ with respect to $\lambda_{max}$. Combined with the inequality (3), we obtain the upper bound of $\sum_k\mathcal{B}_n(\vA^{_{(k)}})$.

\textbf{Example 4}. Consider an $n+1$-partite generalized GHZ state \cite{GHZ}:
\begin{eqnarray}
\ket{\Phi}_{\vA}=\cos\theta\ket{0}^{\otimes n+1}+\sin\theta\ket{1}^{\otimes n+1},
\end{eqnarray}
where $\theta\in (0,\frac{\pi}{2})$. Define $M^{a_1^0}, \cdots, M^{a_{n+1}^0}\in \{\sin\theta\ket{0}+\cos\theta\ket{1}, \cos\theta\ket{0}-\sin\theta\ket{1}\}$. For each outcome $a_k^0=0$, the resultant is an $n$-partite maximally entangled GHZ state. This implies for some quantum correlations derived from local measurement \cite{Sy} that
\begin{eqnarray}
\sum_{k=1}^{n+1}\mathcal{B}_n= (n+1)2^{n-1}\sqrt{2}> (n+2)2^{n-1}
\label{SD7}
\end{eqnarray}
which violates the inequality (\ref{npartite}) for any $\theta\in (0,\frac{\pi}{2})$. This has verified the genuine multipartite nonlocality of any GHZ state. Similar result holds for high-dimensional GHZ states.

\section{Hierarchic correlations from networks}

 Consider a quantum state prepared using a set of entangled states. This allows to extend the entanglement theory to network scenarios that can be applied for quantum network communication \cite{Yang2022,Jiang2024}. Given a network configuration, it imposes special restrictions on the shared sources for each party \cite{BGP,GMTR,Luo2018,Supic2020,Contreras,TPLR}, i.e., each outcome depends on all the states that come from certain sources. Inspired by recent results \cite{Alex,LuoMX2023,TPLR}, we present hierarchic network nonlocalities.

\begin{figure}
\begin{center}
\includegraphics[width=0.45\textwidth]{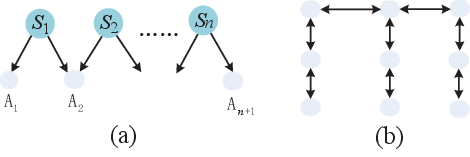}
\end{center}
\caption{\small (Color online) (a) Multipartite chain network. Each two adjacent parties $\sA_i$ and $\sA_{i+1}$ share one bipartite source $S_i$, $i=1, \cdots, n$. (b) Acyclic network. Here, each pair of two parties are connected with one subnetwork consisting of all bipartite sources.}
\label{fig-s4}
\end{figure}

Consider an $n$-partite Bell test with $m$ sources $S_1,\cdots, S_m$ that distributes states to space-like separated observers $\sA_1, \cdots, \sA_n$. The measurement outcome $a_i$ of $\sA_i$ depends on local shares and the type of measurement $x_i$. The joint distribution of all outcomes, conditional on measurement settings, is denoted as $P(\va|\vx)$. Different from  single entanglement, the main goal here is to feature the new nonlocality from networks under the independent assumption of sources. As these sources may be classical variables, quantum states, or NS sources. This leads to hierarchic network correlations as follows:

\begin{definition}
The joint correlation from a given networks is classified as
\begin{itemize}
\item{} $k$-level network nonlocal ($k$-NN) if it cannot be simulated by any probabilistic mixture of networks each of them is constructed by replacing at most $k-1$ quantum states with classical variables while others are replaced by NS sources. The $1$-NN correlation is named as full network nonlocal (FNN) while the $m-1$-NN correlation is network nonlocal (NN).

\item{} $k$-level genuine quantum network nonlocal ($k$-QNN) if it cannot be simulated by any probabilistic mixture of networks each of them is constructed by replacing at most $k-1$ multipartite quantum states with biseparable NS sources. The $1$-QNN correlation is named as genuine quantum network nonlocal (GQNN) while the $m$-GQNN correlation is quantum network nonlocal (QNN).

\item{} $k$-level Bell network nonlocal ($k$-BNN) if it cannot be simulated by any probabilistic mixture of networks each of them is constructed by replacing at most $k-1$ quantum states with classical variables. The $1$-BNN correlation is named a genuine Bell network nonlocal (GBNN) while the $m$-BNN correlation is Bell network nonlocal (BNN).
\end{itemize}

\label{def-1}
\end{definition}

Consider an $n$-partite quantum network consisting of states $\rho_1,\cdots, \rho_m$, where each state $\rho_i$ is at most $n-1$-partite entangled state. Each party $\sA_i$ shares some states of $\rho_j$'s with others. From Definition 2, BNN is the weakest nonlocality for a quantum network. In fact, the BNN correlation does not admit the following decomposition
\begin{equation}
    P(\va|\vx)
    =\int_{\Omega}\prod_{i=1}^{n} P(a_i|x_i,\bar{\lambda}_{i})\prod_{j=1}^md\mu(\lambda_j),
    \label{eqn-E2}
\end{equation}
where $\lambda_j$ denotes the classical variable shared by all parties who share the state $\rho_j$, and $\bar{\lambda}_i$ denotes the set of variables shared by the party $\sA_i$. This kind of network nonlocality may be simulated by using partially entangled sources. Instead, the GFNN is to rule out all correlations from any hybrid realizations of a given network with at least one classical variable and lots of no-signaling sources. This provides the strong nonlocalities \cite{Alex}. The present network nonlocality is different from the recent model \cite{LuoMX2023} by allowing any probabilistic mixture of hybrid realizations of a given quantum network.

To show the main idea, we take the $n+1$-partite chain network shown in Fig.\ref{fig-s4}(a). Here, each two parties $\sA_i$ and $\sA_{i+1}$ share one bipartite source $S_i$, $i=1, \cdots, n$. For any network consisting of at least $k$ classical variables and other NS sources, similar to Lemma 1 all the correlations of $\sA_1$ and $\sA_{n+1}$ satisfy the CHSH inequality \cite{CHSH}. This implies the same bound for any mixed hybrid networks. For a quantum realization consisting of generalized EPR states $|\phi_i\rangle:=\cos\theta_i\ket{00}+\sin\theta_i\ket{11}$ with $\theta_i\in (0, \frac{\pi}{2})$, $i=1,\cdots, n$. There are local projection measurements for all parties $\sA_2, \cdots, \sA_{n}$, such that they can activate an EPR state for $\sA_1$ and $\sA_{n+1}$. This implies the maximal violation of the CHSH inequality with some quantum correlations of $\sA_1$ and $\sA_{n+1}$. This has verified the multipartite correlations with the lifting method. For verifying hierarchic correlations from chain networks we consider the following inequalities
\begin{eqnarray}
&&\sum_{i=1}^{n-1}\textsf{CHSH}(A_i^{(i)},A_{i+2}^{(i)})
\nonumber\\
&\leq&
\left\{
\begin{array}{lll}
4n-4, & \textsf{NS};
\\
4k-8+2\sqrt{2}(n-k+1), & \textsf{kNSQ};
\\
2\sqrt{2}(n-1), & \textsf{Q};
\\
2n+2k-6, & \textsf{kNSC};
\\
2\sqrt{2}(k-2)+2(n-k+1), & \textsf{kQC};
\\
2n-2, & \textsf{C},
\end{array}
\right.
\label{trip}
\end{eqnarray}
where \textsf{kNSQ} denotes correlations from a chain network consisting of at most $k$ NS bipartite sources and others being quantum states, \textsf{Q} denotes quantum correlations from a quantum chain network, \textsf{kNSC} denotes correlations from a chain network consisting of at most $k$ NS bipartite sources and others being classical variables, \textsf{kQC} denotes correlations from a chain network consisting of at most $k$ bipartite quantum states and others being variables, and \textsf{C} denotes classical  correlations. The proof is easy for correlations from a chain network consisting of all NS sources, quantum states, or classical variables. Moreover, for a tripartite network consisting of a NS source and quantum state shared by $\sA_i,\sA_{i+1},\sA_{i+2}$, similar to Lemma 1, the joint correlation of $\sA_i$ and $\sA_{i+2}$ can maximally violate the CHSH inequality \cite{SI}. This implies the result for \textsf{kNSQ}. For a tripartite network consisting of a NS source and classical variable shared by $\sA_i,\sA_{i+1}$ and $\sA_{i+2}$, the joint correlations satisfy the CHSH inequality \cite{SI}. This implies the result for \textsf{kNSC} or \textsf{kQC}.

In general, consider an $n$-partite connected acyclic network $\cN$ consisting of sources $S_1, \cdots, S_N$, $n\geq 3$. An example is shown in Fig.\ref{fig-s4}(b).  In this network, each pair of two parties are connected by a unique subnetwork consisting of some independent sources.  Denote $\hat{\sA}_1, \cdots, \hat{\sA}_t$ as all parties who are independent, i.e., they do not share any source with each other. Suppose that any hybrid network consists of at least $N-t+1$ classical variables and other NS sources. All the parties $\hat{\sA}_1, \cdots, \hat{\sA}_k$ share at least one classical variable with other parties. Similar to Lemma 1, the following Svetlichny inequality for all parties $\hat{\sA}_1, \cdots, \hat{\sA}_k$ conditional on other's local measurements holds as \cite{Sy}:
 \begin{equation}
  \mathcal{B}_{k}(\tilde{\sA}_1, \cdots, \tilde{\sA}_k)\leq 2^{k-1}.
   \label{chain2}
\end{equation}
The same bound applies to any combination of these hybrid networks.

In a quantum network comprising generalized EPR states or generalized GHZ states, local projection measurements \cite{ES,Luo2023} can be performed for all parties $\tilde{\sA}_1, \cdots, \tilde{\sA}_k$, leading to the collapsed state of $\hat{\sA}_1, \cdots, \hat{\sA}_k$ being a maximally entangled GHZ state. This indicates that there exist quantum correlations capable of maximally violating the Svetlichny inequality \eqref{chain2} with certain quantum correlations of $\hat{\sA}_1, \cdots, \hat{\sA}_k$, such as the $N-k+1$-NN and $N-k+1$-GNN. Specifically, for an $n$-partite star network involving $\sA_1,\cdots, \sA_{n}$ and $\sB$, we can examine its GFNN and GQNN. In this setup, each pair of parties $\sA_i$ and $\sB$ shares a bipartite source, necessitating the design of specific sets of Bell inequalities to assess hierarchical correlations in general networks.

\begin{figure}
\begin{center}
\includegraphics[width=0.4\textwidth]{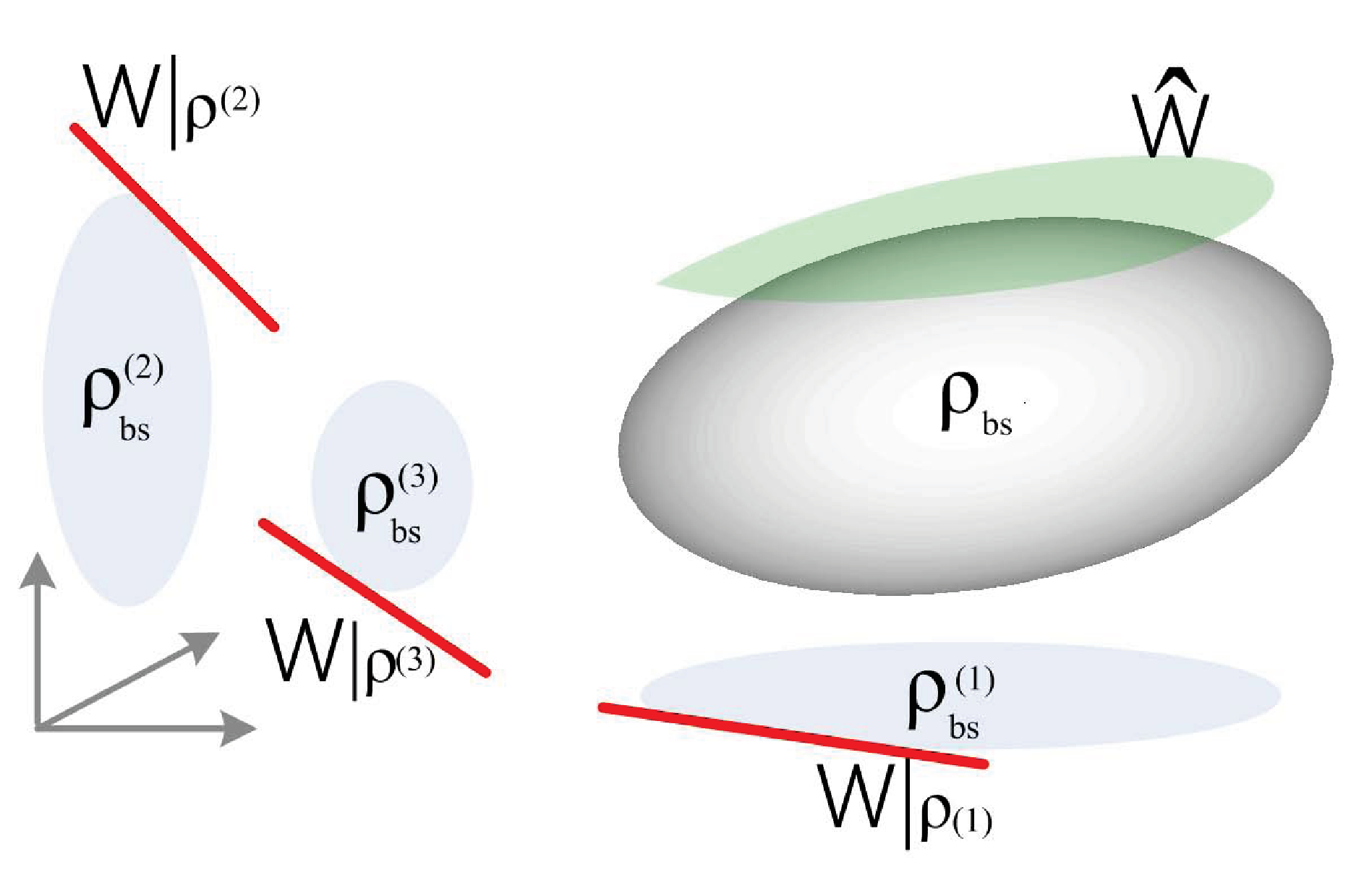}
\end{center}
\caption{\small (Color online) Lifting an entanglement witness $\cW$ on $n$-partite states to design a witness operator (\ref{entangleW}). Here, $\rho^{_{(i)}}_{bs}$ denotes $n$-partite biseparable states without the particle $i$.
}
\label{fig-5}
\end{figure}

\section{Genuine multipartite entanglement witness}

Consider a linear operator to witness genuine $n$-partite entanglement as $\cW$, it satisfies ${\rm Tr}(\rho_{bs}\cW)\leq 0$ for any biseparable state $\rho_{bs}$ and ${\rm Tr}(\rho\cW)>0$ for some entangled state $\rho$ \cite{HHH}. Define an $n+1$-partite operator as
\begin{eqnarray}
\hat{\cW}:=\sum_{i=1}^{n+1}\cW|_{\rho^{_{(i)}}}-c_q,
\label{entangleW}
\end{eqnarray}
where $c_{q}$ denotes the upper bound of $\cW$ in terms of quantum state, and $\rho^{_{(i)}}$ is the resultant conditional on the local measurement of one particle. $\hat{\cW}$ is an $n+1$-partite entanglement witness for verifying genuine entanglement. In fact, consider a biseparable state on Hilbert space $\otimes_{j=1}^{n+1}\cH_{A_j}$ as
\begin{eqnarray}
\rho_{bs}=\sum_{i=1}^{n+1}p_{i}\rho_{A_i}\otimes \rho_{\vA_i},
\label{eqn-F1}
\end{eqnarray}
where $\vA_{i}$ denotes all parties except for $\textsf{A}_i$, and $\rho_{\vA_i}$ denote the joint state. Similar to Lemma 1, the final state of any observer $\textsf{A}_i$, is given by $
\rho^{_{(i)}}=p_i\varrho^{_{(i)}}+(1-p_i)\varrho_{bs}^{_{(i)}}$,  where $\varrho^{_{(i)}}$ may be any quantum state and  $\varrho_{bs}^{_{(i)}}$ is a biseparable state. Similar to the inequality (\ref{npartite}) it follows that
\begin{eqnarray}
\sum_{i=1}^{n+1}\cW|_{\rho^{_{(i)}}}=
\sum_{i=1}^{n+1}p_i\cW|_{\varrho^{_{(i)}}}+\sum_{i=1}^{n+1}(1-p_i)\cW|_{\varrho_{bs}^{_{(i)}}} \leq  c_q
\label{eqn-F3}
\end{eqnarray}
by using the inequalities of $\cW|_{\varrho^{_{(i)}}}\leq c_q$ and $\cW|_{\varrho_{bs}^{_{(i)}}}\leq 0$. This entanglement witness is useful for verifying genuine multipartite entanglement by lifting local entanglement witness, as shown in Fig.\ref{fig-5}.

Take an $n+1$-qubit GHZ state \cite{GHZ} as an example. We have $\cW=(n-1)\mathbbm{1}-\sum_{k=1}^{n}S_{k}$ \cite{Toth2005}, where the maximal quantum bound of $\hat{\cW}$ is $n$ with $c_q=1$, and the operators $S_1=\sigma_x^{\otimes n}, S_k=\sigma_z^{(k-1)} \otimes \sigma_z^{(k)}$ on the $k-1$-th and $k$-th qubits for $k=2,\cdots, n$. This allows to witness entanglement with two Pauli measurements and one projection measurement beyond the projector-based witness with four measurement settings \cite{GH}. This can be extended to verify cluster and graph states \cite{Hein}.

\section{Discussion}

The unified method enables the construction of other Bell inequalities to assess hierarchical correlations (Section E \cite{SI}). These distinct levels of nonlocal correlations offer varying degrees of security for different applications. Quantum secret sharing and quantum multi-party secure computing demand the strongest GMN$^*$ correlations to defend against joint attacks by certain participants \cite{Hillery,Cleve,Moreno,Luo2022}. GMN/GFNN correlations can ensure security in quantum conference key distribution scenarios \cite{HJP2020,DBW,LH}. In the realm of quantum multi-party blind computation utilizing resource states \cite{RUV,Li2014,Luo2022}, GMN may guarantee the integrity and confidentiality of computational tasks, while GEPRN is sufficient for quantum distributed computation \cite{CE1999}. Quantum state verification may necessitate GME for a specific state \cite{DY,BA,Luo2022a}. These tasks typically involve multipartite correlations. Intriguingly, certain quantum game tasks \cite{Mermin,Luo2019} may demonstrate quantum supremacy solely with MBN.

Moreover, the lifting method is robust against noise. As for the generalized GHZ state with white noise \cite{Werner}, the inequality (\ref{tripartite}) provides a unified noise visibility of $4/3\sqrt{2}$ for verifying GMN beyond Svetlichny inequality \cite{Sy,Mermin} or Hardy inequality \cite{Chen}. This result can be easily extended to multipartite GHZ states \cite{GHZ} by using the inequality (\ref{npartite}). It allows for a smaller noise visibility of $1/\sqrt{2}$ for verifying GMN$^{(2)}$ of noisy GHZ or W states. Similar results can be extended to noisy quantum networks. These results may highlight further investigations on new approaches for verifying the quantum correlations of many-body systems or quantum networks.

\section*{Supporting Information}

Supporting Information is available from the Wiley Online Library.

\section*{Acknowledgements}

This work was supported by the National Natural Science Foundation of China (Nos.62172341,12204386, 12075159, and 12171044), Sichuan Natural Science Foundation (No.2023NSFSC0447), Beijing Natural Science Foundation (No.Z190005), Interdisciplinary Research of Southwest Jiaotong University China Interdisciplinary Research of Southwest Jiaotong University China (No.2682022KJ004), and the specific research fund of the Innovation Platform for Academicians of Hainan Province (No. YSPTZX202215).

\section*{Conflict of Interest}

The authors declare no other conflict of interest.

\section*{Author Contributions}

M.X.L. conducted the research. All authors wrote and reviewed the manuscript.

\section*{Data Availability Statement}

This is no data generated in research.

\appendix

\section*{A. The biseparable no-signaling model}

The biseparable no-signaling (NS) model is based on a general NS source, rather than classical variables or quantum states, as the source \cite{PR,BLM,Barrett}. In order to ensure non-negative statistics, this model generally satisfies specific hypotheses \cite{PR,BLM,Barrett} including \textit{Convexity}, \textit{Distinguishability}, \textit{Commutativity}, \textit{Statistics}, \textit{NS principle}, and \textit{Linearity}.

Especially, the state space $\{\dt\}$ and measurement space $\{M_a\}$ are both convex,  which can be understood through the probabilistic mixture in classical physics. Different measurements can be distinguished such as $M_a,N_b$ for outcomes $a$ and $b$,  respectively. Local operations on distinct subsystems commute, and the statistics of measurement outcomes satisfy the classical distribution. For composite systems, the joint distribution satisfies the no-signaling (NS) principle \cite{PR}. Additionally, the joint distribution is a linear function of local measurements and the involved state. All of these features are relevant to the measurement statistics that follow.

Consider a general biseparable probability distribution \cite{Sy}:
\begin{eqnarray}
P_{bs}(\va|\vx)
=\sum_{I,\oI}p_{I,\oI}P(\va_I|\vx_I) P(\va_{\oI}|\vx_{\oI}),
\label{biseparablen}
\end{eqnarray}
where $P(\va_I|\vx_I)$ is the joint distribution of the outcomes $\va_I$ conditional on the measurement settings $\vx_I$, and similar notation is used for $P(\va_{\oI}|\vx_{\oI})$. Both $\{P(\va_I|\vx_I)\}$ and $\{P(\va_{\oI}|\vx_{\oI})\}$ may  any NS distributions. The linearity assumption allows us to generate the biseparable no-signaling (NS) source using an $n$-partite statistical test, as described in \cite{Luo2023}. The source is given by
\begin{eqnarray}
\dt_{A_1\cdots{}A_n}:=\sum_{I,\oI}p_{I,\oI}\dt_{I}\bot\dt_{\oI}.
\label{NSsourcen}
\end{eqnarray}
where $\dt_{I}$ denotes the NS state of all systems $A_i$ with $i\in I$, $\dt_{\oI}$ is the NS state of all systems $A_j$ with $j\in \oI$, and $\bot$ indicates that the two NS states are independent of each other. This formulation provides a method for characterizing the biseparable correlations in subsequent discussions.

\section*{B. Proof of Lemma 1}
\label{subDa}

\begin{center}
\begin{figure*}
\includegraphics[width=0.95\textwidth]{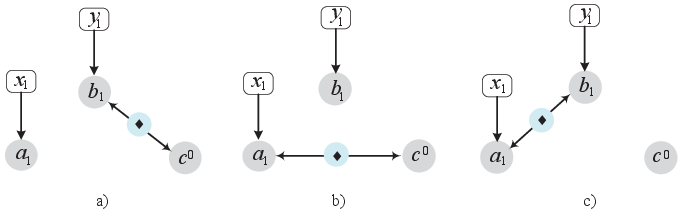}
\caption{\small (Color online) Schematic causal networks of a Bell-type test with biseparable source. Observers $\sA,\sB$ and $\sC$ share the source (\ref{NSsourcen}). }
\label{fig-6}
\end{figure*}
\end{center}

Consider an $n+1$-partite Bell-type test using the biseparable NS source (\ref{NSsourcen}), see one example shown in Fig.\ref{fig-6}. From the general hypothesizes there is a non-negative mapping $F$ such that
\begin{eqnarray}
P(\va_{k}^{_{(k)}},a_k^0|\vx_{k}^{_{(k)}})=F(\dt,M^{a_k^0}, N_{a_s^{_{(k)}}|x_s^{_{(k)}}}, s\not=k),
\label{SA2}
\end{eqnarray}
where $\dt$ denotes the total state. This distribution $\{P(\va_{k}^{_{(k)}},a_k^0|\vx_{k}^{_{(k)}})\}$  satisfies all the statistical properties similar to classical probability distributions. The joint distribution (\ref{SA2}) is independent of the underlying physical implementation of a biseparable NS source (\ref{NSsourcen}). Note that the biseparable NS source (\ref{NSsourcen}) can be regarded as a classical mixture of $n+1$ separable NS source $\dt_{A_i}\bot \dt_{\vA_{i}}$ under the classical distribution $\{p_i\}$. From the linearity hypothesis, it is equivalent the following two subcases. One is the separable NS source $\dt_{A_k}\bot \dt_{\vA_{k}}$. The other is separable NS sources $\dt_{A_s}\bot \dt_{\vA_{s}}$ with $s\not=k$. (a) Suppose that all observers share a separable NS source $\dt_{A_k}\bot \dt_{\vA_{k}}$, where all observers in $\vsA_k$ share an NS source $\dt_{\vA_k}$. From the NS principle and the independence of the systems owned by two sets $\{\sA_k\}$ and $\{\vsA_k\}$, Eq.(\ref{SA2}) implies a joint distribution as:
\begin{eqnarray}
P^{k)}(\va_{k}^{_{(k)}},a_k^0|\vx_{k}^{_{(k)}})=p^{k)}(a_k^0)P^{k)}(\va_k^{_{(k)}}|\vx_k^{_{(k)}}),
\label{SA3}
\end{eqnarray}
where $p^{k)}(a_k^0)$ is the distribution of the outcome $a_k^0$, and $P^{k)}(\va_k^{_{(k)}}|\vx_k^{_{(k)}})$ is the joint distribution of the outcomes $\va_k^{_{(k)}}$ conditional on the measurement setting $\vx_k^{_{(k)}}$. (b) Suppose that all observers share a separable NS source $\dt_{A_s}\bot \dt_{\vA_{s}}$ for $s\not=k$. Here, all observers in $\vsA_s$ share an NS source $\dt_{\vA_s}$. From the NS principle and the independence of the systems owned by two sets $\{\sA_s\}$ and $\{\vsA_s\}$, the joint distribution (\ref{SA2}) allows the following decomposition:
\begin{eqnarray}
P^{s)}(\va_k^{_{(k)}},a_k^0|\vx_{k}^{_{(k)}})=P^{s)}(a_s^{_{(k)}}|x_s^{_{(k)}})p^{s)}(\va_{sk}^{_{(k)}},a_k^0
|\vx_{sk}^{_{(k)}}),
\label{SA4}
\end{eqnarray}
where $\va_{sk}^{_{(k)}}=(a_j^{_{(k)}},j\not=s,k)$, $\vx_{sk}^{_{(k)}}=(x_j^{_{(k)}},j\not=s,k)$, $p^{s)}(a_s^{_{(k)}}|x_s^{_{(k)}})$ is the distribution of the outcomes $a_s^{_{(k)}}$ conditional on the measurement setting $x_s^{_{(k)}}$, and $P^{s)}(\va_{sk}^{_{(k)}},a_k^0|\vx_{sk}^{_{(k)}})$ is the joint distribution of the outcome $\va_{sk}^{_{(k)}}$ and $a_k^0$, conditional on the measurement setting $\vx_{sk}^{_{(k)}}$.

In summary, for any outcomes of $(\va_k^{_{(k)}},a_k^0)$, by the linearity hypothesis with the source (\ref{NSsourcen}), Eqs.(\ref{SA3}) and (\ref{SA4}) imply a joint distribution given by
\begin{eqnarray}
P(\va_k^{_{(k)}},a_k^0|\vx_{k}^{_{(k)}})
&=&\sum_{s\not=k}p_sp^{s)}(a_s^{_{(k)}}|x_s^{_{(k)}})P^{s)}(\va_{sk}^{_{(k)}},a_k^0|\vx_{sk}^{_{(k)}})
\nonumber
\\
&&+p_kp^{k)}(a_k^0)P^{k)}(\va_k^{_{(k)}}|\vx_k^{_{(k)}}).
\label{SA5}
\end{eqnarray}

From the composite hypothesis the distribution of the outcome $a_k^0$ is given by
\begin{eqnarray}
p(a_k^0)=\sum_{\va_k^{_{(k)}}}P(\va_k^{_{(k)}},a_k^0|\vx_{k}^{_{(k)}})
\label{SA6}
\end{eqnarray}
for any given setting $\vx_k^{_{(k)}}$. According to the Bayes' rule, Eqs.(\ref{SA5}) and (\ref{SA6}) imply a conditional distribution given by
\begin{eqnarray}
P(\va_k^{_{(k)}}|\vx_{k}^{_{(k)}};a_k^0)&:=&\frac{P(\va_k^{_{(k)}},a_k^0|\vx_{k}^{_{(k)}})}{p(a_k^0)}
\nonumber
\\
&=&\sum_{s\not=k}p_s\frac{p^{s)}(a_s^{_{(k)}}|x_s^{_{(k)}})
P^{s)}(\va_{sk}^{_{(k)}},a_k^0
|\vx_{sk}^{_{(k)}})}{p(a_k^0)}
+p_k\frac{p^{k)}(a_k^0)
 P^{k)}(\va_k^{_{(k)}}|\vx_k^{_{(k)}})}{p(a_k^0)}
\nonumber
\\
&=&\sum_{s\not=k}p_sp^{s)}(a_s^{_{(k)}}|x_s^{_{(k)}}) \frac{P^{s)}(\va_{sk}^{_{(k)}},a_k^0|\vx_{sk}^{_{(k)}})}{p(a_k^0)}
+p_k  P^{k)}( \va_k^{_{(k)}}|\vx_k^{_{(k)}})
\nonumber\\
&:=&p_k  P_{ns}(\va_k^{_{(k)}}|\vx_k^{_{(k)}})+(1-p_k)P_{bs}(\va_k^{_{(k)}}|\vx_{k}^{_{(k)}};a_k^0)
\label{SA7}
\end{eqnarray}
for $p(c^0)\not=0$, where $P_{ns}(\va_k^{_{(k)}}|\vx_k^{_{(k)}})$ is a conditional distribution defined by $P_{ns}(\va_k^{_{(k)}}|\vx_k^{_{(k)}}):=p^{k)}(\va_k^{_{(k)}}|\vx_k^{_{(k)}})$, and $P_{bs}(\va_k^{_{(k)}}|\vx_{k}^{_{(k)}};a_k^0)$ is a conditional distribution defined by $P_{bs}(\va_k^{_{(k)}}|\vx_{k}^{_{(k)}};a_k^0):=\sum_{s\not=k}p_sp^{s)}(a_s^{_{(k)}}|x_s^{_{(k)}}) P^{s)}(\va_{sk}^{_{(k)}},a_k^0|\vx_{sk}^{_{(k)}})/p(a_k^0)
+p_k  P^{k)}(\va_k^{_{(k)}}|\vx_k^{_{(k)}})$.

From Eqs.(\ref{SA4}-\ref{SA7}), the distribution $P_{ns}(\va_k^{_{(k)}}|\vx_k^{_{(k)}})$ can be a general NS distribution. The joint distributions $p^{s)}(a_s^{_{(k)}}|x_s^{_{(k)}}) P^{s)}(\va_{sk}^{_{(k)}},a_k^0|\vx_{sk}^{_{(k)}})/p(a_k^0)$ for all $s\not=k$ are bipartite separable in terms of $a_s^{_{(k)}}$ and $\va_{sk}^{_{(k)}}$. This implies that the distribution $P_{bs}(\va_k^{_{(k)}}|\vx_{k}^{_{(k)}};a_k^0)$ is bipartite separable \cite{Bell}.  This has proved Lemma 1. $\Box$

\section*{C. Proof of Inequality (7)}

Take a Bell experiment as shown in Figure 1 in the main text, with a biseparable NS source (\ref{NSsourcen}), where observers $\sA_1$ and $\sA_2$ share an NS source $\dt_{A_1A_2}$ with probability $p_3$, observers $\sA_1$ and $\sA_3$ share an NS source $\dt_{A_1A_3}$ with a probability $p_2$, and observers $\sA_2$ and $\sA_3$ share an NS source $\dt_{A_2A_3}$ with a probability $p_1$.

\textit{1) Biseparable NS source}

The proof of the inequality (7) in the main text is followed using Lemma 1 and the CHSH inequality \cite{CHSH}, where $c_{ns}=4$. Especially, from the linearity of CHSH operator \cite{CHSH} and the inequality (3) in the main text it follows that
\begin{eqnarray}
\sum_{k=1}^{3}\mathsf{CHSH}(P(\va_k^{_{(k)}}|\vx_{k}^{_{(k)}};a_k^0))
&\leq & 4(p_1+p_2+p_3)+ 2(3-p_1-p_2-p_3)\leq 8
\label{SC1}
\end{eqnarray}
from $\sum_{i=1}^3p_i=1$, for any $k=1, 2, 3$.

\textit{2) Biseparable quantum source}

Consider a tripartite Bell experiment consisting of biseparable quantum correlations
$\dt^q_{A_1A_2A_3}=p_1\dt_{A_1}\bot\dt^q_{A_2A_3}+ p_2\dt_{A_2}\bot\dt^q_{A_1A_3}+ p_3\dt^q_{A_1A_2}\bot\dt_{A_3}$, where $\sA_1$ and $\sA_2$ share a quantum state $\dt^q_{A_1A_2}$ with probability $p_3$, $\sA_1$ and $\sA_3$ share a quantum state $\dt^q_{A_1A_3}$ with probability $p_2$, and $\sA_2$ and $\sA_3$ share a quantum state $\dt^q_{A_2A_3}$ with probability $p_1$. $\dt_{X}$ denotes the local source of $X\in \{A_1,A_2,A_3\}$.

Similar to the proof of Lemma 1, the joint distribution of the outcomes of $\sA_s$ and $\sA_t$, conditional on the outcomes $a_k^0$ of the observer $\sA_k$, is given by $P_{a_k^0}(\va_{k}^{_{(k)}}|\vx_k^{_{(k)}})=p_kP_{q}(\va_k^{_{(k)}}|\vx_k^{_{(k)}})
+(1-p_k)P_s(\va_k^{_{(k)}}|\vx_k^{_{(k)}})$, where $P_{q}(\va_k^{_{(k)}}|\vx_k^{_{(k)}})$ can be any quantum distribution and $P_s(\va_k^{_{(k)}}|\vx_k^{_{(k)}})$ is separable. Combining with the linearity of CHSH operator \cite{CHSH} we get that
\begin{eqnarray}
\sum_{k=1}^3\textsf{CHSH}(P_{a_k^0}(\va_{k}^{_{(k)}}|\vx_k^{_{(k)}}))
&\leq & 2\sqrt{2} (p_1+p_2+p_3)+2 (3-p_1-p_2-p_3)\leq 4+2\sqrt{2}
\label{SC4}
\end{eqnarray}
from $\sum_{i=1}^3p_i=1$, for any $k=1, 2, 3$.

\textit{3) Fully separable source}

Consider a fully separable source $\dt_{A_1A_2A_3}=\dt_{A_1}\bot\dt_{A_2}\bot\dt_{A_3}$. It is easy to show that the joint distribution of the outcomes of $\sA_s$ and $\sA_t$, conditional on the outcomes $a_k^0$ of $\sA_k$, is separable. This implies that
\begin{eqnarray}
\sum_{k=1}^3\textsf{CHSH}(P_{a_k^0}(\va_{k}^{_{(k)}}|\vx_k^{_{(k)}}))
\leq 6
\label{SC6}
\end{eqnarray}
from the CHSH inequality \cite{CHSH}.

\textit{4) Other sources}

The quantum bound of $6\sqrt{2}$ can be derived from CHSH inequality. For a no-signaling source it implies the upper bound of $12$ for any no-signaling sources.

\textbf{Example S1}. Consider a triangle network consisting of three generalized EPR states $|\phi_i\rangle=\cos\theta_i\ket{00}+\sin\theta_i\ket{11}$ on Hilbert space $\cH_{2i-1}\otimes \cH_{2i}$ with $\theta_i\in (0, \frac{\pi}{2})$, $i=1, 2, 3$. Here, $\sA_1$ has particles $1$ and $6$, $\sA_2$ has particles $2$ and $3$, and $\sA_3$ has particles $4$ and $5$. Define $M^{a_1^0}\in \{\sin\varphi_1\ket{00}+\cos\varphi_i\ket{11}, \cos\varphi_1\ket{00}-\sin\varphi_1\ket{11}, \ket{01},\ket{10}\}$ with $\tan\varphi_1=1/(\tan\theta_1\tan\theta_2)$, $M^{a_2^0}\in \{\sin\varphi_2\ket{00}+\cos\varphi_i\ket{11}, \cos\varphi_2\ket{00}-\sin\varphi_2\ket{11}, \ket{01},\ket{10}\}$ with $\tan\varphi_2=1/(\tan\theta_1\tan\theta_2)$. $M^{a_3^0}\in \{\sin\varphi_3\ket{00}+\cos\varphi_3\ket{11}, \cos\varphi_3\ket{00}-\sin\varphi_3\ket{11}, \ket{01},\ket{10}\}$ with $\tan\varphi_3=1/(\tan\theta_2\tan\theta_3)$. $M_{a_2}^{_{(1)}}$, $M_{a_1}^{_{(2)}}$, $M_{a_1}^{_{(3)}}  \in \{\sigma_z\otimes \mathbbm{1},\sigma_x\otimes \mathbbm{1}\}$ and $M_{a_3}^{_{(1)}}, M_{a_3}^{_{(2)}},  M_{a_2}^{_{(3)}} \in \{(\sigma_z\pm \sigma_x)/\sqrt{2}\otimes \mathbbm{1}\}$. For the outcome $a_1^0=0$ of $\textsf{A}_1$, $\sA_2$ and $\sA_3$ share an EPR state and one generalized EPR state. This implies the maximal violation of CHSH inequality for some local measurements of $\sA_2$ and $\sA_3$. The same result holds for some joint correlations of $\sA_1$ and $\sA_2$ or $\sA_1$ and $\sA_3$, i.e., $P_{a_2^0}(a_1^{_{(2)}}, a_3^{_{(2)}}|x_1^{_{(2)}},x_3^{_{(2)}})$and $P_{a_3^0}(a_1^{_{(3)}}, a_2^{_{(3)}}|x_1^{_{(3)}},x_2^{_{(3)}})$. So, the quantum correlations derived from a triangle network achieve the maximal quantum bound of $6\sqrt{2}$ for any $\theta_i\in (0,\frac{\pi}{2})$. This shows the genuine tripartite nonlocality of any triangle network by violating the inequality (\ref{SC1}), where $a_i$ satisfy $a_i,a_j\not=0$ for at least two different integers $i,j$ and $\sum_i a_i^2=1$. This can be extended for high-dimensional generalized EPR states.

\section*{D. Proof of Inequality (19)}

\textit{1) Biseparable NS source}

The proof of the inequality (19) in the main text is followed using Lemma 1 and the Svetlichny-Mermin inequality \cite{Sy}. Here,  $\mathcal{B}_n$ denotes the $n$-party Svetlichny-Mermin-type operator that can be defined recursively as:
\begin{equation}
  \mathcal{B}_n=(A_{x_n=0}+A_{x_n=1})B_{n-1}+(A_{x_n=0}-A_{x_n=1})B_{n-1}'
  \label{SyMer}
\end{equation}
with $B_1=A_{x_1=0}$, and $B_k'$ being obtained from $B_k$ by flipping all the inputs of all the parties. For the Svetlichny-Mermin-type inequality, we have $c_{ns}=2^n$. From the linearity of Svetlichny-Mermin-type operator, it follows that
\begin{eqnarray}
\sum_{k=1}^{n+1}\mathcal{B}_n(P(\va_k^{_{(k)}}|\vx_{k}^{_{(k)}};a_k^0)) \leq 2^n+ n2^{n-1} = (n+2)2^{n-1}.
\label{SD2}
\end{eqnarray}

\textit{2) Biseparable quantum source}

Consider an $n+1$-partite Bell experiment consisting of biseparable quantum source $\dt^q_{\vA}=\sum_{k=1}^{n+1}p_k\dt_{A_k}\bot\dt^q_{\vA_k}$ where $\vsA_k$ share a quantum state $\dt^q_{\vA_k}$ with probability $p_k$, $k=1, \cdots, n+1$. $\dt_{X}$ denotes local state of $X\in \{A_1,\cdots, A_{n+1}\}$.

Similar to the proof of Lemma 1, the joint distribution of the outcomes of $\vsA_k$, conditional on the outcomes $a_k^0$ of $\sA_k$, is given by
\begin{eqnarray}
P_{a_k^0}(\va_{k}^{_{(k)}}|\vx_k^{_{(k)}})=p_kP_{q}(\va_k^{_{(k)}}|\vx_k^{_{(k)}})
+(1-p_k)P_s(\va_k^{_{(k)}}|\vx_k^{_{(k)}}),
\label{SD4}
\end{eqnarray}
where $P_{q}(\va_k^{_{(k)}}|\vx_k^{_{(k)}})$ may be any $n$-partite quantum distribution and $P_s(\va_k^{_{(k)}}|\vx_k^{_{(k)}})$ is a bipartite separable quantum distribution. Note that the quantum bound for the Svetlichny-Mermin operator is given by $2^{n-1}\sqrt{2}$. From the linearity of the Svetlichny-Mermin operator, we get that
\begin{eqnarray}
\sum_{k=1}^{n+1}\mathcal{B}_n(P_{a_k^0}(\va_{k}^{_{(k)}}|\vx_k^{_{(k)}}))
&\leq & \sum_{k=1}^{n+1}p_k 2^{n-1}\sqrt{2}+ (n+1-\sum_{k=1}^{n+1}p_k) 2^{n-1}
\nonumber
\\
&=& (n+\sqrt{2})2^{n-1}.
\label{SD5}
\end{eqnarray}

\textit{3) Fully separable source}

Consider a fully separable source $\dt_{\vA}=\dt_{A_1}\bot\cdots\bot \dt_{A_{n+1}}$. The joint distribution of the outcomes of $\vsA_k$, conditional on the outcomes $a_k^0$ of the observer $\sA_k$, is fully separable. This implies that
\begin{eqnarray}
\sum_{k=1}^{n+1}\mathcal{B}_n(P_{a_k^0}(\va_{k}^{_{(k)}}|\vx_k^{_{(k)}}))
\leq (n+1)2^{n-1}.
\label{SD6}
\end{eqnarray}

\textit{4) Other sources}

The quantum bound of $(n\sqrt{2}+\sqrt{2})2^{n-1}$ is followed from Svetlichny-Mermin inequality. For a no-signaling source, we have the maximal bound of $2^n$ for Svetlichny-Mermin operator. This implies the upper bound of $(n+1)2^{n}$ for any no-signaling sources.

\subsection*{2. Genuine multipartite nonlocality}

\begin{figure}
\begin{center}
\includegraphics[width=0.6\textwidth]{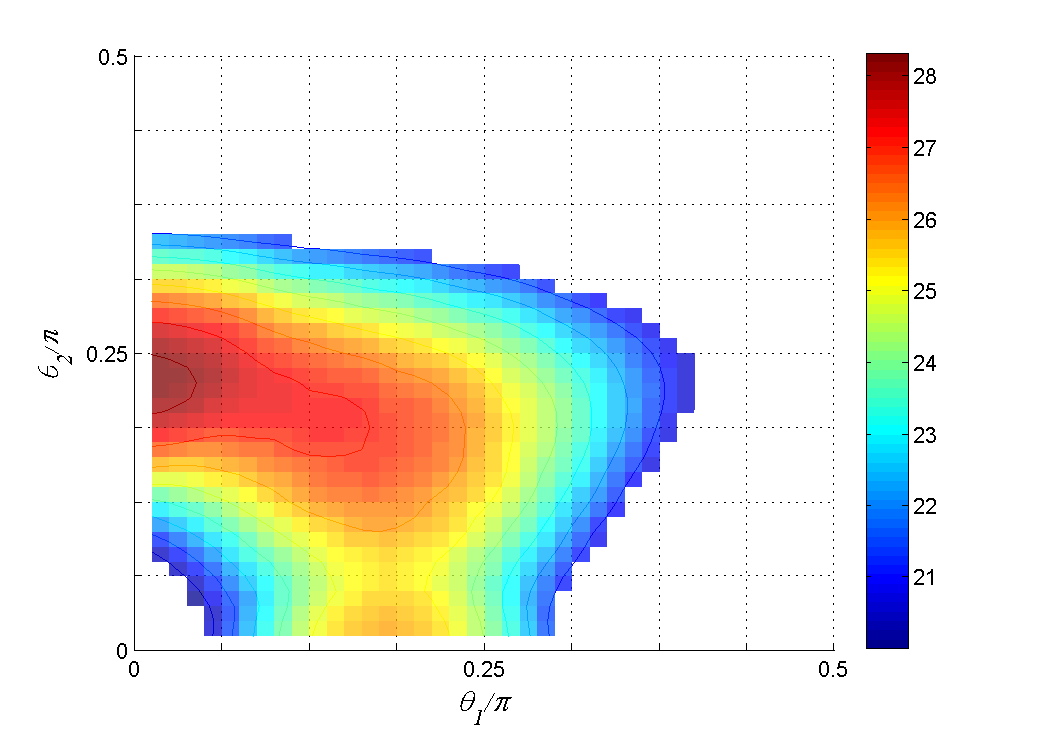}
\end{center}
\caption{\small (Color online) Upper bound (\ref{S50}) in terms of generalized W states (\ref{WW}). Here, $\alpha_1=\cos\theta_1\cos\theta_2, \alpha_3=\cos\theta_1\sin\theta_2$ and $\alpha_4=\sin\theta_1$.}
\label{FigSW}
\end{figure}

\textbf{Example S2}. Consider a generalized W state \cite{Dur}:
\begin{eqnarray}
|W\rangle=\sum_{k=1}^{n+1}a_k\ket{1}_k,
\end{eqnarray}
where $a_k\not=0, \sum_{k}a_k^2=1$, where $\ket{1}_k$ denotes the $k$-th qubit is in the state $|1\rangle$ while all the others are in $|0\rangle$. Define $M^{a_1^0}, \cdots, M^{a_{n+1}^0}\in \{\ket{0},\ket{1}\}$. For each measurement outcome $a_k^0=0$, the resultant is an entangled $n$-partite W state. The total bound depends on the maximal bound of each $n$-partite Mermin-Svetlichny-type operator in terms of generalized W state \cite{Dur}. Take the four-qubit W state
\begin{eqnarray}
|W_4\rangle=\alpha_1\ket{0001}+\alpha_2\ket{0010}+\alpha_3\ket{0100}+\alpha_4\ket{0100},
\label{WW}
\end{eqnarray}
where $a_i$ satisfy $\sum_i\alpha_i^2=1$. Define the local measurement $M^{a_i^0}\in \{|0\rangle, |1\rangle\}$. For the outcome $a_1=0$, there is a generalized tripartite W state $|W_3\rangle=\alpha_1\ket{001}+\alpha_2\ket{010}+\alpha_3\ket{100}$. From the result \cite{AR} we get that
\begin{eqnarray}
\max_{A,B,C}\mathcal{B}_3(A,B,C)&=&-\sin \Theta + \sin \Theta_a + \sin \Theta_b + \sin \Theta_c
\nonumber\\	
& &+ C_{13}(\sin \Theta + \sin \Theta_a -\sin \Theta_b + \sin \Theta_c)
\nonumber\\	
& &+ C_{12}(\sin \Theta - \sin \Theta_a + \sin \Theta_b + \sin \Theta_c)
\nonumber\\	
& &+ C_{23}(\sin \Theta + \sin \Theta_a + \sin \Theta_b - \sin \Theta_c),
\end{eqnarray}
where $\Theta=\theta_1+\theta_2+\theta_3$, $\Theta_x=\Theta-2\theta_x$, $\theta_i$ are measurement parameters, and $C_{ij}=2\alpha_i\alpha_j$. The total bound is then given by
\begin{eqnarray}
\sum_{i=1}^4\mathcal{B}_3(P_{a_i^0}(\va_i^{_{(1)}}|\vx_i^{_{(1)}}))
&=&\sum_{i=1}^4[-\sin \Theta^{_{(i)}} + \sin \Theta_a^{_{(1)}} + \sin \Theta_b^{_{(1)}} + \sin \Theta_c^{_{(i)}}
 \nonumber\\	
& & + C_{13}^{_{(i)}}(\sin \Theta^{_{(i)}} + \sin \Theta_a^{_{(i)}} - \sin \Theta_b^{_{(i)}} + \sin \Theta_c^{_{(i)}})
\nonumber\\	
& &+ C_{12}^{_{(i)}}(\sin\Theta^{_{(i)}} - \sin \Theta_a^{_{(i)}} + \sin \Theta_b^{_{(i)}} + \sin \Theta_c)
 \nonumber\\	
& &  + C_{23}^{_{(i)}}(\sin \Theta^{_{(i)}} + \sin \Theta_a^{_{(i)}} + \sin \Theta_b^{_{(i)}} - \sin \Theta_c^{_{(i)}})],
\end{eqnarray}
where $\Theta^{_{(i)}}=\theta_1^{_{(i)}}+\theta_2^{_{(i)}}+\theta_3^{_{(i)}}$, $\Theta_x^{_{(i)}}=\Theta^{_{(i)}}-2\theta_x^{_{(i)}}$, $\theta_j^{_{(i)}}$ are measurement parameters, $C_{12}^{_{(1)}}=C_{12}^{_{(2)}}=2\alpha_1\alpha_2$, $C_{13}^{_{(2)}}=C_{13}^{_{(3)}}=2\alpha_1\alpha_4$, $C_{23}^{_{(2)}}=C_{13}^{_{(4)}}=2\alpha_2\alpha_4$, $C_{12}^{_{(3)}}=C_{13}^{_{(1)}}=2\alpha_1\alpha_3$, $C_{23}^{_{(3)}}=C_{23}^{_{(4)}}=2\alpha_3\alpha_4$, and $C_{23}^{_{(1)}}=C_{12}^{_{(4)}}=2\alpha_2\alpha_3$. For special case of $\alpha_1=\alpha_2$, $\theta_k^{_{(i)}}=\theta_k^{_{(j)}}$ for any $k=1,2,3$ and $i\not=j$. It follows that
\begin{eqnarray}
\sum_{i=1}^4\mathcal{B}_3(P_{a_i^0}(\va_i^{_{(1)}}|\vx_i^{_{(1)}}))
&=&4(-\sin \Theta + \sin \Theta_a + \sin \Theta_b + \sin \Theta_c)
 \nonumber\\	
& & + 2(3\alpha_1\alpha_4+\alpha_1^2)(\sin \Theta+ \sin \Theta_a- \sin \Theta_b + \sin \Theta_c)
\nonumber\\	
& &+2(3\alpha_1\alpha_3+\alpha_1^2)(\sin\Theta - \sin \Theta_a+ \sin \Theta_b + \sin \Theta_c)
 \nonumber\\	
& &  + 2(2\alpha_3\alpha_4+\alpha_2\alpha_4+\alpha_2\alpha_3)(\sin \Theta+ \sin \Theta_a + \sin \Theta_b- \sin \Theta_c).
\label{S50}
\end{eqnarray}
The numeric evaluations are shown in Fig.\ref{FigSW}.

\textbf{Example S3}. Consider an $n+1$-partite chain network as shown in Figure 3(a), consisting of $n$ generalized EPR states $|\phi_i\rangle=\cos\theta_i\ket{00}+\sin\theta_i\ket{11}$ on Hilbert space $\mathcal{H}_{2i-1}\otimes \mathcal{H}_{2i}$ with $\theta_i\in (0, \frac{\pi}{2})$, $i=1, \cdots, n$. Here, $\sA_1$ and $\sA_{n+1}$ has  particle $1, 2n$, respectively, and $\sA_i$ has particles $2i$ and $2i+1$. One method is first to transform the $n+1$-partite chain network into an $n+1$-partite GHZ state under the local operations and classical communication. This is reasonable by regarding local operations as pre-processing of nonlocal game. It allows us to verify all the genuine multipartite nonlocality from Example S6.

Another method is without pre-processing. Define $M^{a_1^0}, M^{a_3^0}\in \{\ket{0},\ket{1}\}$, $M^{a_2^0}\in \{\sin\varphi\ket{00}+\cos\varphi\ket{11}, \cos\varphi\ket{00}-\sin\varphi\ket{11}, \ket{01},\ket{10}\}$ with $\tan\varphi=1/(\tan\theta_1\tan\theta_2)$, $M_{a_2}^{_{(1)}},M_{a_1}^{_{(2)}}, M_{a_1}^{_{(3)}} \in \{\sigma_z,\sigma_x\}$, and $M_{a_3}^{_{(1)}}\in \{\cos\hat{\theta}_1\sigma_z\pm \sin\hat{\theta}_1\sigma_x\}$, $M_{a_3}^{_{(2)}}\in \{\cos\hat{\theta}_2\sigma_z\pm \sin\hat{\theta}_2\sigma_x\}$ and $M_{a_2}^{_{(3)}}\in \{\cos\hat{\theta}_3\sigma_z\pm \sin\hat{\theta}_3\sigma_x\}$. From the
maximal violation of CHSH inequality is given by $2\sqrt{1+\sin^2\theta_1}$ for $\sA_2$ and $\sA_3$, $\sqrt{1+\sin^2\theta_2}$ for $\sA_1$ and $\sA_2$, or $2\sqrt{2}$ for $\sA_1$ and $\sA_3$  (conditional on other's outcomes). This implies the following inequality
\begin{eqnarray}
\sum_{k=1}^3\textsf{CHSH}(P_{a_k^0}(\va_k^{_{(k)}}|\vx_k^{_{(k)}}))
=2\sqrt{2}+2\sqrt{1+\sin^2\theta_1}+2\sqrt{1+\sin^2\theta_2}.
\end{eqnarray}

\textbf{Example S4}. Consider an $n+1$-partite complete-connected network consisting of $(n+1)n/2$ generalized EPR states $|\phi_i\rangle:=\cos\theta_i\ket{00}+\sin\theta_i\ket{11}$ on Hilbert space $\cH_{2i-1}\otimes \cH_{2i}$ with $\theta_i\in (0, \frac{\pi}{2})$, $i=1,\cdots, (n+1)n/2$. Here, each pair of two parties $\sA_i$ and $\sA_j$ share one bipartite entanglement. Define $M^{a_i^0}\in \{\sin\varphi_i\ket{0}^{\otimes n}+\cos\varphi_i\ket{1}^{\otimes n}, \cos\varphi_i\ket{0}^{\otimes n}-\sin\varphi_i\ket{1}^{\otimes n}, \ket{\vec{i}},\forall \vec{i}\not=\vec{0}, \vec{1}\}$ with $\tan\varphi_i=1/\prod_{j\in S_i}\tan\theta_j$ and $\vec{i}=i_1\cdots i_n$, where $S_j$ denotes the set of integer $i$ satisfying that $\sA_i$ shares one EPR state $|\phi_s\rangle$ with others for any $s\in S_i$. For the outcome $a_i^0=0$ of $\textsf{A}_i$, $\vsA_{i}$ share one $n$-partite GHZ state $(\ket{0}^{\otimes n}+\ket{1}^{\otimes n})/\sqrt{2}$. This implies that the quantum correlations from proper local measurements maximally violate the Svetlichny-Mermin inequality as
\begin{eqnarray}
\mathcal{B}_n(P_{a_i^0}(\va_i^{_{(1)}}|\vx_i^{_{(1)}}))
= (n+1)2^{n-1}\sqrt{2}> (n+2)2^{n-1}.
\end{eqnarray}
So, the quantum correlations from an $n+1$-partite complete-connected network can achieve the maximal quantum bound of $(n+1)2^{n-1}\sqrt{2}$ for any $\theta_i\in (0,\frac{\pi}{2})$.

\section*{E. Related to other inequalities }

The Svetlichny inequality is useless to verify the following tripartite entangled state $|\Phi\rangle=\sqrt{3}/2\ket{000}+\sqrt{3}/4\ket{110}+1/4\ket{111}$  \cite{Bancal2013}. Similar to Example S2 we get the maximal quantum bound $2\sqrt{2}+4\sqrt{1.75}>8$, which violates the inequality (\ref{SC1}).

Another method is using the facet inequality \cite{Bancal2013}. In general, it follows new inequalities for verifying the 4-partite genuine nonlocality as
\begin{eqnarray}
\sum_{i=1}^{4}\mathcal{B}|_{\{P_i(\va_i|\vx_i,a_i^0)\}}\leq c_{ns}+3c
\label{SE1}
\end{eqnarray}
for any tripartite facet inequality $\mathcal{B}(A,B,C)\leq c$ \cite{Bancal2013}, where $c_{ns}$ denotes the NS bound.

One example is given by
\begin{eqnarray}
\sum_{i=1}^{4}\mathcal{B}|_{\{P_i(\va_i|\vx_i,a_i^0)\}}\leq 0,
\end{eqnarray}
where $\mathcal{B}=-2P(A_1B_1)-2P(B_1C_1)-2P(A_1C_1)-P(A_0B_0C_1)-P(A_0B_1C_0)-P(A_1B_0C_0) +2P(A_1B_1C_0)+2P(A_1B_0C_1)+2P(A_0B_1C_1)+2P(A_1B_1C_1)$, $P(A_iB_j ):=P(a=0,b=0|x=i,y=j)$, and $P(A_iB_jC_k):=P(a=0,b=0,c=0|x=i,y=j,z=k)$. This can be further lifted for verifying multipartite genuine nonlocality.

The other method is by extending the Hardy inequality \cite{Hardy,Chen2014}. This implies new chain inequality for verifying genuine multipartite correlations as
\begin{eqnarray}
\sum_{k=1}^{n+1}\mathcal{B}(P(\va_k^{_{(k)}}|\vx_{k}^{_{(k)}};a_k^0))
\leq 0,
\label{SE2}
\end{eqnarray}
where $\mathcal{B}$ is defined as
$\mathcal{B}:=P(\vec{0}|\vx)-\sum_{k}P(\vec{0}|x_k,\vx'_{k})-
\frac{1}{n-1}\sum_{k\not=k'}P(1_k,1_{k'},\vec{0}_{kk'}|x_{k},x_{k'},\vx'_{kk'})$ \cite{Chen2014} with two different measurement settings $x_i$ and $x_i'$ for each party.

\end{document}